\documentclass[runningheads]{llncs}
\usepackage{graphicx}
\usepackage{svg}
\svgsetup{inkscapelatex=false}
\usepackage{appendix}
\usepackage{booktabs}
\usepackage{amsmath}
\usepackage[misc,geometry]{ifsym}
\usepackage{tikz,xcolor,hyperref}
\newcommand{\orcid}[1]{\href{https://orcid.org/#1}{\includegraphics[width=10pt]{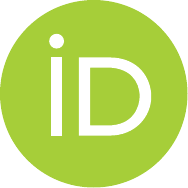}}}

\begin{document}
\title{New Perspectives on Sensitivity and Identifiability Analysis using the Unscented Kalman Filter}
\author{Harry Saxton\inst{1}$^{\textrm{(\Letter)}}$\orcid{0000-0001-7433-6154} \and
Xu Xu\inst{2} \orcid{0000-0002-9721-9054} \and
Ian Halliday\inst{3} \and 
Torsten Schenkel\inst{2} \orcid{0000-0001-5560-1872}}
%ZZ
\authorrunning{H. Saxton et al.}

\institute{Materials \& Engineering Research Institute, Sheffield Hallam University, Howard Street, Sheffield, S1 1WB, United Kingdom \\
\email{H.saxton@shu.ac.uk} \and
Department of Engineering and Mathematics, Sheffield Hallam University, Howard Street, Sheffield, S1 1WB, United Kingdom \\ \and
Department of Infection, Immunity and Cardiovascular Disease, University of Sheffield, The Medical School, Beech Hill Road, Sheffield, S10 2RX, United Kingdom}
\maketitle              % typeset the header of the contribution
\begin{abstract}
Detailed dynamical systems’ models used in the life sciences may include hundreds of state variables and many input parameters, often with physical meaning. Therefore, efficient and unique input parameter identification, from experimental data, is an essential but challenging task for this class of model. To clarify our understating of the process (which within a clinical context amounts to a personalisation), we utilise the computational methods of Unscented Kalman filtration (UKF), sensitivity and orthogonality analysis. We have applied these three techniques to a test-bench model of a single ventricle, coupled, via Ohmic valves, to a Compliance-Resistor-Compliance (CRC) Windkessel electrical analogue model of the systemic circulation, chosen in view of its relative simplicity, interpretability and prior art. Utilising an efficient, novel and real-time implementation of the UKF\footnote{Code available at https://github.com/H-Sax/CMSB-2023}, we show how, counter-intuitively, input parameters are efficiently recovered from experimental data \emph{even if they are not sensitive parameters in the currently accepted sense}. This result (i) exposes potential limitations in the standard interpretation of what it means for an input parameter to be designated \emph{identifiable} and (ii) suggests that the concepts of sensitivity and identifiability may have a weaker relationship than commonly thought - at least in the presence of an appropriate data set. We rationalise these observations. 

Practically, we present results which show the UKF to be an efficient method for assigning personalised input parameters from experimental data \emph{in real-time}, which enhances the clinical significance of our approach.  

\end{abstract}
\section{Introduction}
Mechanistic mathematical models of the cardiovascular system (CVS) can be made 
representative of  physiology and pathophysiology \cite{hose2019cardiovascular} - 
provided they have accurately optimised input parameters, ideally from 
suitable patient data \cite{dash2022non,colunga2023parameter}. The relationship between model functionality and available data must be carefully balanced - a parsimonious model with sufficient functionality, deployment protocols, the properties of the target data and physiology are all to be considered. 
  
Three standard approaches are used for the development of CVS models: (i) complex three-dimensional (3D) models with many degrees of freedom \cite{quarteroni2016geometric} to provide local insight, (ii) one-dimensional (1D) models in which only the streamwise flow co-ordinate and time are retained, and which filter information to reach spatially extended transient physiology such as pulse wave dynamics and (iii) zero-dimensional (0D) lumped parameter, electrical analogue or compartmental models \cite{shi2011review}, with the ability to simulate the dynamics of the extended CVS. Closed-loop 0D models are, typically, carefully curated combinations of low-order sub-models of hemodynamic effects. They are expressed in terms of passive electrical analogues - resistors representing mechanical dissipation in flow, capacitors organ compliance and inductors any flow inertial effects. We are concerned exclusively with such 0D models.
  
Each compartment of a 0D model represents physiology at a level which is often clinically amenable, with its corresponding input parameters then serving as potential bio-markers with clinical significance \cite{huberts2018needed,pironet2019practical}.
To exploit this, we must deduce methodologies which efficiently assign optimal personalised parameters preferably from patient data (which correspond to a model output), by inverse operation of the CVS model. This process, termed model personalisation, or parameter identification, therefore assumes a central significance in the calibration of any $0$D model to (patho)physiology in the individual and, by extension, to clinical deployment. This process is not exclusive to CVS models of course; parameter identification algorithms are vital for the deployment of almost all biological systems' models\cite{bekey1978identification,simpson2022parameter,gabor2015robust}. 
 
 The task of CVS $0$D model personalisation is, apparently, a practical exercise in model input parameter identification. When considering model input parameter identifiability, one is concerned with two essential types: structural, and practical \cite{guillaume2019introductory} (a weak case, accounting for, e.g., the practical consequences of experimental error in the target data). Here, we concern ourselves with practical identifiability only. Given the paucity and notorious inconsistency of clinical data, the obvious practical response to the problem of model personalisation is to characterise the patient by a co-ordinate in input parameter space - a model operating point - or (more likely) a finite but bounded region, or a subspace.
But, how does one discover such a subspace?
 
Canonically, model personalisation is  rooted in assessing the input parameters' sensitivity against given output metrics. Sensitivity analysis (SA) studies how a change in a model’s output can be apportioned to different sources of uncertainty among its likely many input parameters \cite{saltelli2008global}. Often, a lack of model sensitivity is deemed the root cause of input parameters’ unidentifiability - when model parameters' variation does not shift outputs, their value cannot be inferred \cite{gul2016mathematical,schiavazzi2017patient,eisenberg2017confidence}. Two types of SA exist: (i) local SA addresses sensitivity relative to change of a 
single parameter value
\emph{one at a time} about a fixed operating point; (ii) global SA examines sensitivity with regard to the entire parameter distribution \cite{marino2008methodology}. For personalised medicine, it is deemed essential to examine the whole parameter space \cite{lazarus2022sensitivity}. SA can expose influential parameters in sets, the members of which all have similar effects across surveyed outputs; without more data, 
it then becomes very challenging to determine each uniquely. Accordingly, if one does not examine -and screen- such dependency between parameters, selecting parameters on influence alone, one does not know if a chosen subset of parameters are all individually identifiable. There are several methods within the literature which combine both ideas to produce the optimal input parameter set \cite{li2004selection,ottesen2014structural,olsen2019parameter}.
 
Another method of ingesting data into our cardiovascular (CV) model uses an Unscented Kalman filter (UKF) \cite{wan2001unscented}. Recently, progressive data-assimilation approaches have increased in popularity, due to quick computation and the UKF's ability fully to exploit time-varying measurements (in contradistinction to the scalar mean, minimum or maximum indices which are often used). 
It has been employed in a range of CV models \cite{lam2017use,liu2022noninvasive,marchesseau2013personalization,canuto2020ensemble}. Within this methodology, identifiable parameters are characterised by a low variance - which is the object of the algorithm. However which of the methods cited above correctly identifies the optimal input parameter? We present a systematic comparison and analysis of typical methods, used to deduce the optimal set of input parameters. Our comparison involves (i) examination of orthogonality within the model, (ii) global sensitivity and (iii) the UKF algorithm. We develop and detail a novel implementation which is real-time and adaptable to any system described by a set of differential equations. 
 
We structure as follows: Section 2 details the methods of and the implementation used within the work; Section 3 details the results and Section 4 gives a comprehensive comparison and discussion of the optimal input parameter sets which are returned by the methods.

\section{Methods}
We describe the model under investigation, the sensitivity analyses performed, our parameter orthogonality analysis and, finally, the implementation of the UKF to the basic single ventricle model. All computation was performed using Julia \cite{bezanson2017julia}, using packages including DifferentialEquations.jl \cite{rackauckas2017differentialequations} and Distributions.jl \cite{JSSv098i16} to solve the differential algebraic equation (DAE) system \cite{wanner1996solving} formulation and implement the UKF.
Specifically, simulations were solved using Rodas5P algorithm \cite{wanner1996solving}, relative and absolute tolerances were set to $10^{-12}$. We enforced a time step of $0.00225$ (444 time steps per cycle) and ran the model for 30 cycles, given a steady solution was reached after 3 cycles. We used Makie.jl to visualise results \cite{DanischKrumbiegel2021}, and GlobalSensitivity.jl to perform the global sensitivity analysis \cite{Dixit2022}.
\subsection{DAE Model \& Measurements}
\begin{figure}[htp!]
    \centering
    \includegraphics[width=0.78\linewidth]{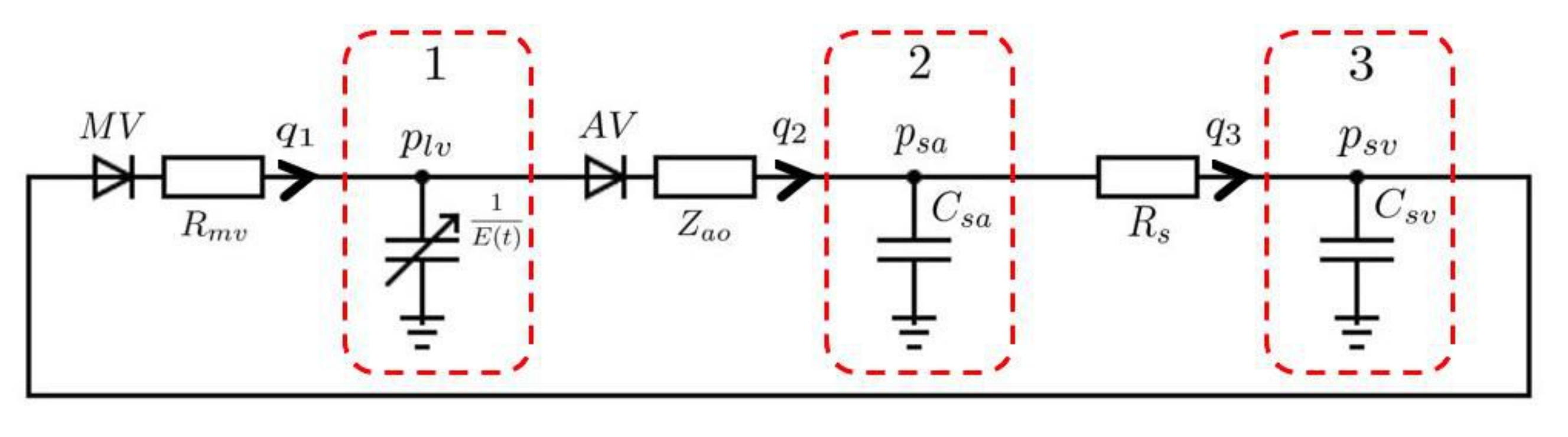}
    \caption{ {\bf The systemic circulation, single ventricle, model.} 
    An electrical analogue representation of our state-space
    system CVS model. 
    The elastance of the left ventricle chamber used is the 
    Shi double cosine \cite{korakianitis2006numerical}. The valves (diodes) 
    are assumed to have Ohmic behaviour, under both forward and reverse bias, with a very large regurgitating resistance. Our textual notations for the resistors and the capacitors are defined in Table \ref{tbl:measurements}.}
    \label{fig:model}
\end{figure}

Our base mechanical model is a three-compartment system-level, DAE based, electrical analogue CV model after Bjordalsbakke et. al. \cite{bjordalsbakke2022parameter}. See Figure \ref{fig:model}. Each compartment state is specified by 
its dynamic pressure $P(t)$ (mmHg), an inlet flow $Q(t)$ (mL/s) and a volume $V(t)$ (mL): 
\begin{align}
\label{Model}
\underline{X}_i(t)=\left(V_i(t),P_i(t),Q_i(t) \right)^{T}, \quad i \in \{lv,sa,sv \}, 
\end{align}
where $\underline{X}$ represent the state variables of of our system, $lv$ denotes the left ventricle, $sa$ the systemic arteries and $sv$ the venous system. Formally, $t$ is a continuous variable. Our system is canonically 
expressed in a compact, state-space form: 
\begin{equation}
\begin{aligned}
\label{State_Space}
\frac{d}{dt} \underline{X}(t) = \underline{f} (\underline{X}(t) &; \underline{\theta}),
\quad \underline{Y}(t) = \underline{h}(\underline{X}(t)), \\
\underline{Y}(t) &= (P_{lv}, P_{sa}, V_{lv})^{T}
\end{aligned}
\end{equation}
in which $\underline{\theta}$ denotes the input parameter set declared in Table \ref{tbl:measurements} as a vector, function $\underline{h}$ is the measurement operator. A full model description, its parameters and model the solution can be found in Appendix \ref{app-modelderiv}.

\begin{table}

\caption{{\bf Model input parameters. } Each input parameter's unit is stated alongside 
a chosen initial value. $\tau$ is the cardiac cycle length and is fixed such that $\tau = 1 s.$\label{tbl:measurements}
}

\begin{tabular}{ |c|c|c|c| } 
\hline
Parameter $\underline{\theta}$ (units) & Description & Initial Value \\
\hline
$E_{max}\ \mathrm{\left[\frac{mmHg}{ml}\right]}$ & Maximal ventricular contractility  & 1.5 \\ 
$E_{min}\ \mathrm{\left[\frac{mmHg}{ml}\right]}$& Minimal ventricular contractility  & 0.03 \\ 
$\tau_{es}\ \mathrm{(s)}$ & End systolic time & 0.3$\tau$ \\ 
$\tau_{ep}\ \mathrm{(s)}$ & End pulse time & 0.45$\tau$ \\ 
$Z_{ao}\ \mathrm{\left[\frac{mmHg\hspace{1mm} s}{ml}\right]}$ & Aortic valve resistance & 0.033 \\ 
$R_{mv}\ \mathrm{\left[\frac{mmHg\hspace{1mm} s}{ml}\right]}$ & Mitral valve resistance & 0.06 \\ 
$R_{s}\ \mathrm{\left[\frac{mmHg\hspace{1mm} s}{ml}\right]}$ & Systemic resistance & 1.11 \\ 
$C_{sa}\ \mathrm{\left[\frac{mmHg\hspace{1mm} s}{ml}\right]}$ & Systemic compliance & 1.13 \\ 
$C_{sv}\ \mathrm{\left[\frac{mmHg\hspace{1mm} s}{ml}\right]}$ & Venous compliance & 11.0 \\
\hline
\end{tabular}
    \centering
\end{table}

To examine SA, orthogonality and the UKF, we derive noisy forward data from numerical solutions based upon nominally true parameter values. We generate waveform data for $lv$ pressure $P_{lv}$, $lv$ volume $V_{lv}$ and systemic pressure $P_{sa}$ for 30 cycles, 
representing a clinical scenario of continuous measurements from (say) echocardiography for $V_{lv}$ \cite{fresiello2015cardiovascular} and arterial line measurement for $P_{sa}$ \cite{saugel2020measure}. (Cardiac cathederisation can be performed to extract $P_{lv}$ \cite{keshavarz2020diagnostic}). To supply surrogates for the metrics defined above, our model's numerical solutions are subject to  multiplicative corruption as follows:
\begin{align}
\label{math_measurements}
    \underline{Y}_{j}^{m} = \underline{h}(\underline{X}(t_{j},\underline{\theta}_{t}), \quad \quad
    \underline{Y}_{j}^{n} = \underline{h}(\underline{X}(t_{j},\underline{\theta}_{t})) \cdot(1+\psi_{j}).
\end{align}
Above, the subscript $j$ denotes sampling time, deemed to be the discrete time of the numerical solution, superscript $n$ indicates a noisy solution and superscript $m$ denotes the measured, un-corrupted numerical solution. $\psi_{j}$ is an independently distributed normal random variable, with zero mean and a standard deviation $0.025$, which is typical \cite{canuto2020ensemble}.  

\subsection{Sensitivity and Orthogonality}
Global SA explores tracts of system input parameter space and decomposes parameters' effects on our representative outputs. Here we consider Sobol analysis \cite{sobol2001global} in a probabilistic framework, which decomposes model outputs' variance into fractions which can be attributed to (sets of) inputs. Often, Sobol analysis is performed on discrete outputs to attribute variance to a specific measurement. Here we examine continuous outputs \cite{alexanderian2020variance} and produce waveform data which demonstrate the sensitivity of each input parameter
\emph{over the cardiac cycle} at each time point. We defer further discussion to Appendix \ref{app-ConSen}.
Rather than average across a time range (which process weights regions of low variance equally to those of high variance) we seek to expose differential sensitivities 
by examining variance-weighted averages:
\begin{equation}
\begin{aligned}
\label{Sobol}
S_{[1,T],i} = \frac{\sum_{k} S_{[1,T],i}\underline{Y}(t_{k})\text{Var}(\underline{Y}(t_{k})}{\sum_{k}\text{Var}(\underline{Y}(t_{k}))}, \\ 
\end{aligned}
\end{equation}
where $S_{1}$ represents the first-order indices which inform on relative inﬂuence of every input (total order indices, $S_{T}$, inform relative influence of every input parameters interactions with others). $\underline{Y}^{m}$ denotes the measured model output and $i$ identifies the particular input parameter whose sensitivity is at issue.

To determine a parameter's net importance across all measurements we use the overall importance measure introduced by Li et al. which uses an eigen-decomposition of the Fisher matrix, to rank input parameters. See \cite{li2004selection} and Appendix \ref{app-ParamImpo}.
 
 We use the measure $d_{ij}$ to measure orthogonality between input parameters $\theta_i$ 
and $\theta_j$.
\begin{equation}
    \begin{aligned}
    \label{Param_orth_indiv}
    d_{ij} = \sin{\bigg[\cos{^{-1}\bigg(\frac{S^{T}_{1,j}\cdot S_{1,i}}{||S_{1,j}||||S_{1,i}||}}}\bigg)\bigg], \quad i,j=1,..,n, \quad d_{ij} \in [0,1].
    \end{aligned}
\end{equation}
Here we concern ourselves only with first order Sobol indices $S_{1}$, which allows us to concentrate solely on a parameter's independent effects. $S_{1,i}$ are multidimensional vectors due to each input parameter, $i/j$, having an independent effect against the 3 measurements explored in this work.

\subsection{Unscented Kalman Filter}
The UKF is made up of two distinct steps, firstly the unscented transform (UT) which is a method for calculating the statistics of a random variable which undergoes a nonlinear transformation \cite{julier2004unscented}. The first distinction is that we assume additive noise through the whole model which is accepted practice for biological systems \cite{liepe2013maximizing,silk2014model,pant2016data}. We generate an augmented vector $\underline{x} = [\underline{X}, \underline{\theta}]$, where $\underline{X}$ and $\underline{\theta}$ are the state variables and input parameters; see (\ref{State_Space}). 
 
We assume that the augmented state vector $\underline{x}$ is a Gaussian random variable (GRV) of dimension $L$ where $L = \text{dim}(\underline{X})+\text{dim}(\underline{\theta})$. Now consider propagating the augmented state-vector through the nonlinear function $\underline{f}$. Here and for most biological systems, the non-linear function is represented by a set of ODEs. We measure $P_{lv},\ P_{sa},\ V_{lv}$, $\underline{Y}(t) = \underline{h}(\underline{x}(t))$, $ \underline{h}$ is the previously-used operator- we acquire the ODE solution to $P_{lv}, \ P_{sa}$ and $V_{lv}$.  Assume our GRV has a mean $\underline{x}_{\mu}$ and a covariance $P_{\underline{x}}$. To compute the statistics on the propagation of our GRV through $\underline{f}$, we construct a matrix $\underline{\underline{\chi}}$ of $2L+1$ sigma vectors $\chi_{i}$, where $i$ represents the $ith$ column of the matrix according to the following. For $t = 0, ..., \infty$:
\begin{equation}
    \begin{aligned}
    \label{Sigma_points}
        \chi_{0,t} = \underline{x}^{A}_{\mu,t}, \hspace{1mm} \chi_{i_{1,t}} = \underline{x}^{A}_{\mu,t}& + \left(\sqrt{(L+\lambda)P^{A}_{\underline{x},t}} \right)_{j}, \hspace{1mm} \chi_{i_{2,t}} = \underline{x}^{A}_{\mu,t} - \left(\sqrt{(L+\lambda)P^{A}_{\underline{x},t}} \right)_{j}, \\
         i_{1} &= j = 1,...,L, \hspace{2mm} i_{2} = L+1, ..., 2L.
    \end{aligned}
\end{equation}
We also compute a set of corresponding weights $W_{i}$:
\begin{equation}
    \begin{aligned}
        W_{0}^{\mu} = \frac{\lambda}{L+\lambda}, \hspace{1mm} W_{0}^{c} &= \frac{\lambda}{L+\lambda} + (1 + \beta - \alpha^{2}),  \hspace{1mm} W_{i}^{\mu} = W_{i}^{c} = \frac{1}{2(L+\lambda)}, \\
        i&=1,...,2L, \hspace{1mm} \lambda = \alpha^{2}(L+\kappa) - L,
    \end{aligned}
\end{equation}
where the superscript $A$ represents the assimilated state and parameter vector, $\lambda$ is a scaling parameter, $\alpha$ determines the spread of sigma points around $\underline{x}_{\mu}$ (we use $\alpha =  10^{-3}$). $\kappa$ is another scaling parameter (here $\kappa=0$). $\beta$ incorporates prior knowledge of which distribution $\underline{x}$ follows, here $\beta=2$ is used as this is optimal for GRV. The matrix square root is performed using a Cholesky decomposition \cite{higham1990analysis} which requires the matrix to be positive definite.  We then propagate each sigma vector through the ODE system such that $\Upsilon_{i} = \underline{f}(\chi_{i})$ and determine the the mean and covariance of $\underline{Y}$ using the weighted sample mean and covariance of the propagated sigma vectors. Before we can do this, we must first define the prediction step in the algorithm 
\begin{equation}
    \begin{aligned}
        \hat{\chi}_{t+1|t} = \underline{f}(\chi_{t}), \hspace{1mm} \Upsilon_{t+1|t} = \underline{h}(\hat{\chi}_{t+1|t}).
    \end{aligned}
\end{equation}
The above have corresponding mean and sample covariance: 
\begin{equation}
    \begin{aligned}
            \underline{x}_{\mu,t+1} = \sum_{i=0}^{2L} W_{i}^{\mu}\hat{\chi}_{i,t+1|t},& \hspace{1mm} P_{\underline{x},t+1} =  \sum_{i=0}^{2L} W_{i}^{c}[\hat{\chi}_{i,t+1|t} - \underline{x}_{\mu,t+1}][\hat{\chi}_{i,t+1|t} - \underline{x}_{\mu,t+1}]^{T} + \delta_{Q} I, 
            \\
            \underline{Y}_{t+1}^{\mu} = \sum_{i=0}^{2L} W_{i}^{\mu}\Upsilon_{i,t+1|t},& \hspace{1mm} P_{\underline{Y},t+1} =  \sum_{i=0}^{2L} W_{i}^{c}[\Upsilon_{i,t+1|t} - \underline{Y}^{\mu,t+1}][\Upsilon_{i,t+1|t} - \underline{Y}^{\mu,t+1}]^{T} + R, 
            \\
            P_{\underline{x}\underline{Y},t+1} = &\sum_{i=0}^{2L} W_{i}^{c}[\hat{\chi}_{i,t+1|t} - \underline{x}_{\mu,t+1}][\Upsilon_{i,t+1|t} - \underline{Y}^{\mu,t+1}]^{T},    
    \end{aligned}
\end{equation}
where $P_{\underline{x}\underline{Y}}$ is designated the cross correlation matrix. $R$ is the additive noise on the predicted measurements. $\delta_{Q} I$ is considered a regularisation term to avoid sigma point collapse \cite{turner2012model,pant2016data}, $I$ is an $L \times L$ identity matrix with $\delta_{Q} =  10^{-8}$.
 
We now correct the prediction that has been made by assimilating the noisy data generated on (\ref{math_measurements}). The Kalman gain matrix is calculated as 
$$
K_{t+1} = P_{\underline{x}\underline{Y},t+1}(P_{\underline{Y},t+1})^{-1},
$$ 
which then leads to:
\begin{equation}
    \begin{aligned}
        \underline{x}^{A}_{\mu,t+1} = \underline{x}_{\mu,t+1} + K_{t+1}(\underline{Y}^{n}_{t+1} - \underline{Y}_{t+1}^{\mu}), 
        \\
        P^{A}_{\underline{x},t+1} = P_{\underline{x},t+1} - K_{t+1}P_{\underline{Y},t+1}K_{t+1}^{T},
    \end{aligned}
\end{equation}
where $\underline{x}^{A}_{\mu,t+1}$ and $P^{A}_{\underline{x},t+1}$ are used to generate new sigma points for the $t+1$ time point. 
 
In order to implement the UKF, we take advantage of the versatile SciML ecosystem within 
Julia. Here we implement a discrete callback which performs the Kalman filtration at 
each time point and returns the corrected result. This has been shown to contribute no 
additional computational time associated with the callback, which explains our ability
to produce the result in real time. Most workers manually discretise the ODE, in order to transform it into a 
discrete time system for the implementation of an UKF. Implementing a callback allows us to take advantage of 
advanced ODE solvers within package DifferentialEquations.jl with improved accuracy. 

\section{Results}
We summarise results from the preliminary global SA and the 
average importance and the orthogonality of input parameters. We 
then examine the results of the Kalman filtration and reinforce this by examining 
relationships within the model equations. 

\subsection{Sensitivity \& Orthogonality Analysis}
\begin{figure}[htp!]
\centerline{\includegraphics[width=\columnwidth]{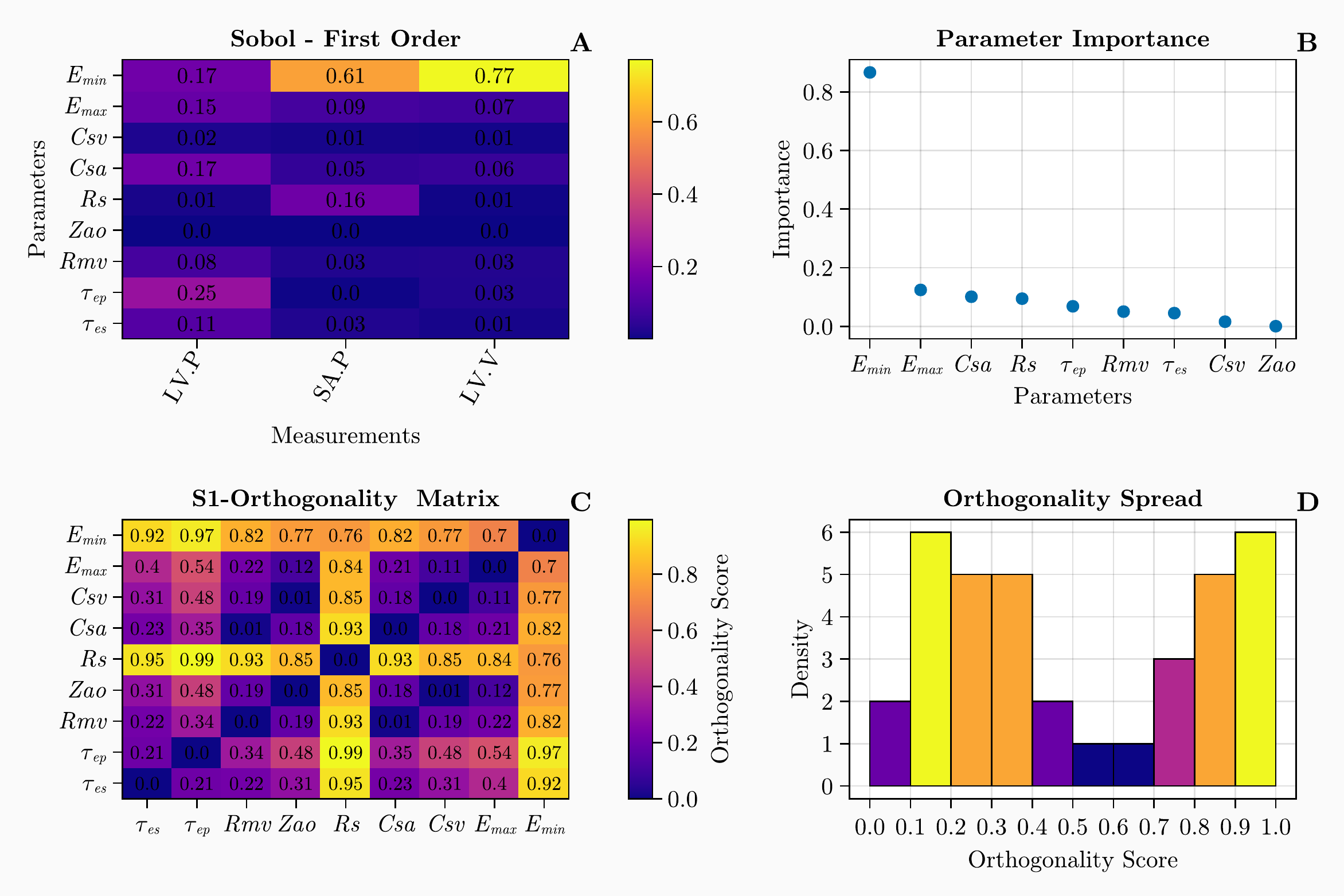}}
\caption{$\boldsymbol{A}$: Sobol indices of ﬁrst order interactions between inputs and outputs; $\boldsymbol{B}$: average parameter importance; $\boldsymbol{C}$: orthogonality score between the Sobol first order matrix; $\boldsymbol{D}$: a histogram showing the orthogonality score spread.}
\label{fig: Sensitivity}
\end{figure}

In Figure \ref{fig: Sensitivity}A, we present the converged Sobol analysis. Each parameter was run for $n=3000$ iterations which, when the first order indices were summed, resulted in values of $0.97,0.98$ and $0.99$ for left ventricular pressure, systemic arterial pressure and left ventricular volume respectively. These results were then checked for $n=5000$ iterations for which we saw no change to 2dp. We see clearly that $E_{min}$ has the strongest influence over all 3 measurements and it appears $R_{s}$ and $\tau_{ep}$ have strong independent influence over arterial pressure and ventricular pressure, respectively. But $C_{sv}$ and $Z_{ao}$ appear to have minimal influence over all 3 measurements. Figure \ref{fig: Sensitivity}B reinforces these findings with $86\%$ of the overall influence concentrated around minimal contractility parameter $E_{min}$, where as only $0.05\%$ of influence an be attributed to the aortic valve resistance $Z_{ao}$. Figure \ref{fig: Sensitivity}C displays the orthogonality of the input parameters in which $0$ denotes total dependence between parameters. We see $E_{min}$ and $R_{s}$ are the most orthogonal parameters in the whole space. It appears as $(C_{sa},R_{mv})$ and $(C_{sv},Z_{ao})$ exhibit the strongest coupling with an orthogonality score of $0.01$, which is surprising. Figure \ref{fig: Sensitivity}D is a histogram showing how the orthogonality appears to concentrated around $[0.1,0.4]$ and $[0.8,1.0)$. This concentration towards high orthogonality scores is due to $R_{s}$ and $E_{min}$ exhibiting high independence over the parameter space. 

\subsection{Kalman Filtration}
\begin{figure}[htp!]
\centerline{\includegraphics[width=\columnwidth]{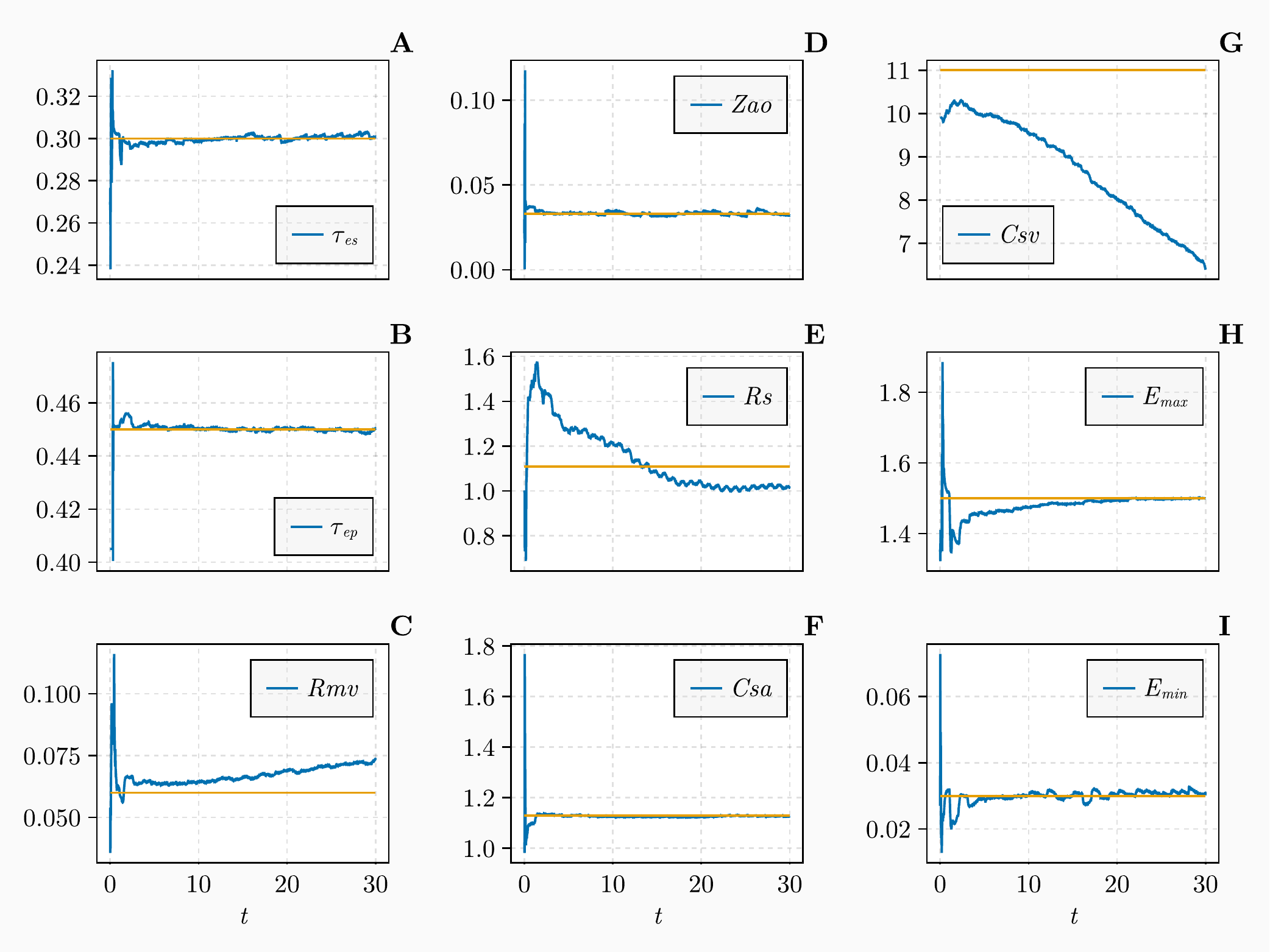}}
\caption{Results of using the UKF to recover input parameter values from noisy synthetic data. The ``true" parameter values are represented as a yellow line, where as the UKF estimate is shown in blue.}
\label{fig: UKF}
\end{figure}

We initialise our parameters $10\%$ from their truth value, while prescribing the largest initial variances to ensure no physiological principles are undermined. From Figure \ref{fig: UKF} we see $\tau_{es}, \tau_{ep}, Z_{ao}, C_{sa}, E_{max}$ and $E_{min}$ are estimated very close to their truth values and that 5 parameters are found within 7 cycles, evidencing considerable facility at recovering truth parameter values. Running the UKF for 30 cycles takes 27.3 seconds which was averaged over 10 successful runs of the UKF. $\tau_{ep}, \tau_{es}$ and $E_{min}$ are the parameters which were found most accurately, with an error of $0.0163\%, \ 0.0607\%$ and $0.339\%$. Then in order of accuracy: $C_{sa}, \ Z_{ao}, \ E_{max}$ all showed error less than $1.3\%$. The parameters with the largest errors were $R_{s}, \ R_{mv}$ and $C_{sv}$ with $2.7\%, \ 12.0\%$ and $21.0\%$ error respectively. We see small fluctuations surrounding the estimation of certain parameters even after the truth value has been found, which relates to the internal dynamics of the system and valve actuation points- where the system reacts to the Heaviside Ohmic valve- and recovers the true parameter value during its current cycle.
 
\begin{figure}[htp!]
\centerline{\includegraphics[width=\columnwidth]{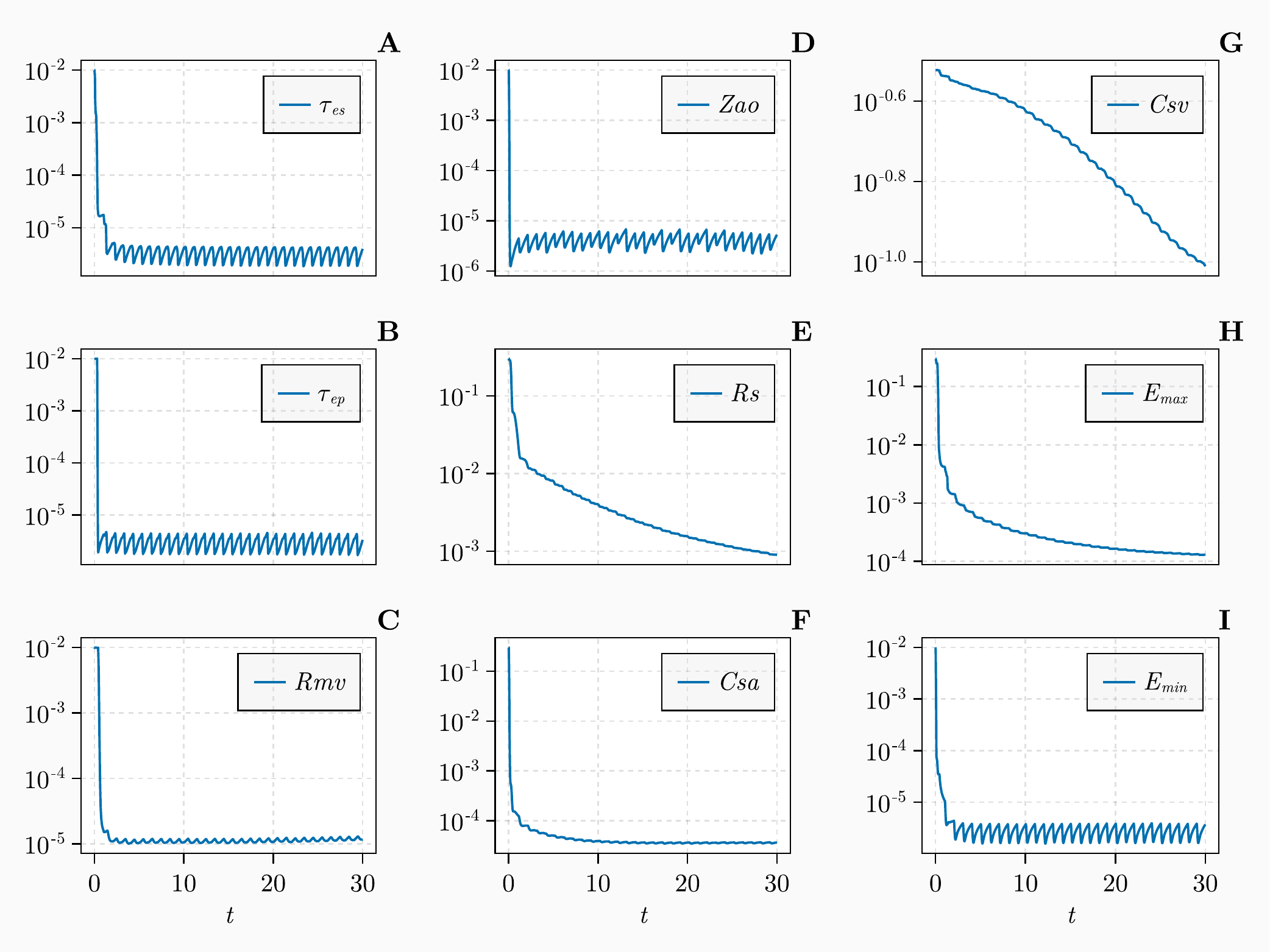}}
\caption{Evolution of the parameter variance during the cardiac cycle. Results are displayed with a log scale.}
\label{fig: UKF_COV}
\end{figure}

Figure \ref{fig: UKF_COV} represents the evolution of the parameter variance as the UKF evolves. We again see the majority of parameters reach a steady variance within 7 cycles, which then oscillates with the cardiac cycle. Parameters with the smallest variance are $\tau_{ep}, E_{min}$ and $\tau_{es}$ with variances $3.36\times 10^{-6}, \ 3.72\times 10^{-6}$ and $3.98\times 10^{-6}$, respectively. The ranked parameters are $Z_{ao}, \ R_{mv}, \ C_{sa}$ which all have a variance less than $1.3\times 10^{-4}$. $E_{max}, \ R_{s}$ and $C_{sv}$ have the largest variances of $1.30\times 10^{-4}, \ 8.96\times 10^{-4}$ and $9.76\times 10^{-2}$. We remark that $R_{mv}$ has low variance, but from the figure we see that it has not settled.

\subsection{Dynamical equations}
Here we derive equations in terms of model input parameters and pressures; volume is derived using $E(t) = \frac{P_{lv}(t)}{V_{lv}(t)}$ \cite{suga1974instantaneous}. 
We break the cycle up into systole and diastole, assuming infinite resistance on the mitral (aortic) valve during systole (diastole). 
For a full derivation see Appendix \ref{app-analytical}. For systole we have 
\begin{equation}
    \begin{aligned}
    \label{Systole}
            \frac{dP_{lv}}{dt} =  \frac{P_{sa} - P_{lv}}{Z_{ao}}E(t) + \frac{P_{lv}}{E(t)}&\frac{dE}{dt}, \hspace{1mm}         \frac{dP_{sa}}{dt} = \frac{P_{lv} - P_{sa}}{C_{sa}Z_{ao}} - \frac{P_{sa} - P_{sv}}{C_{sa}R_s},\\
            \frac{dV_{lv}}{dt}  = \frac{P_{lv} - P_{sa}}{Z_{ao}}, \hspace{1mm} \frac{dP_{sv}}{dt} &= \frac{P_{sa} - P_{sv}}{C_{sv}R_s},
    \end{aligned}
\end{equation}
and for diastole 
\begin{equation}
    \begin{aligned}
        \frac{dP_{lv}}{dt}=  (\frac{P_{sv} - P_{lv}}{R_{mv}})E(t) + &\frac{P_{lv}}{E(t)}\frac{dE}{dt}, \hspace{1mm}  \frac{dP_{sa}}{dt} = \frac{P_{sv} - P_{sa}}{C_{sa}R_s}, \\
        \frac{dP_{sv}}{dt} &= \frac{P_{lv}-P_{sv}}{C_{sv}R_{mv}} + \frac{P_{sa} - P_{sv}}{C_{sa}R_s}.
    \end{aligned}
\end{equation}
From these equations we infer the following:
\begin{enumerate}
    \item $\tau_{es}, \ \tau_{ep}, \ E_{min}, \ E_{max}$ interact directly and independently with with measurements $P_{lv},\ P_{sa},\ V_{lv}$. As derived from the pressure volume relationship specified at the start of section 3.3.
    \item $Z_{ao}$ interacts independently with $P_{lv}, \ P_{sa}, \ V_{lv}$ from ventricular pressure volume equation in (\ref{Systole}).
    \item $C_{sa}$ interacts collectively with $Z_{ao}$ and $R_{s}$ on the measurements $P_{lv}$ and $P_{sa}$ from arterial pressure equation in (\ref{Systole}). $C_{sa}$ directly controls our arterial pressure.
\end{enumerate}
All the above parameters are found with an error of less than $1.5\%$, which we observe due to their independent interactions with measurements. 
We are left with 3 parameters which have an error larger than $2\%$:
\begin{enumerate}
    \item $R_{s}$ interacts collectively with $C_{sa}$ and $C_{sv}$, on the arterial pressure measurement, therefore has no independent interaction with any measured data. $R_{s}$ has an error of $2.7 \%$ and ranks 4th on the importance scale.
    \item $R_{mv}$ interacts with $E(t)$ on $P_{lv}$ and with $C_{sv}$ on $P_{lv}$. Here we see no independent interaction with any measurement. $R_{mv}$ has an error of $12\%$ and ranks $6$th on the importance scale.
    \item $C_{sv}$ interacts with $R_{s}$ on $P_{sa}$ and $R_{mv}$ on $P_{lv}$. Despite $C_{sv}$ interacting with two measurements both it's interactions are conflated by parameters which themselves do not interact independently with any measured experimental data. 
\end{enumerate}
We remark that the parameters which are exposed to both pressure and volume measurements independently are recovered accurately, despite them not being particularly sensitive, 
whereas parameters which are not recovered efficiently often interact collectively with other parameters, so independent effects can not be extracted from the measurements. 

\section{Discussion}

\begin{table}[htp!]

\caption{{\bf Parameter results. } We tabulate the results of the input parameters, ranked as 1 - 9, where, 1st represents the lowest variance, largest importance, largest orthogonality and smallest error. The first rank average (column 5) denotes rank average, displays the parameter's rank excluding error. The second rank average (column 7) includes the error of the UKF estimate. \dag: The error increases to $7.52\%$ given we examine $t>15s$. *: The error decreases to $0.291\%$ given we examine $t>15s$.}
\label{tbl:results}

\small
\begin{tabular}{|c|c|c|c|c|c|c|}
\hline
Parameter &
  \begin{tabular}[c]{@{}l@{}}Variance\\ \& Rank\end{tabular} &
  \begin{tabular}[c]{@{}l@{}}Importance\\ \& Rank\end{tabular} &
  \begin{tabular}[c]{@{}l@{}}Orthog.\\ \& Rank\end{tabular} &
  \begin{tabular}[c]{@{}l@{}}Average \\ Rank\end{tabular} &
  \begin{tabular}[c]{@{}l@{}}Error \%  \\ \& Rank\end{tabular} &  
  \begin{tabular}[c]{@{}l@{}}Average\\ Rank\end{tabular} \\[1mm] \hline
$\tau_{es}$ & $(3.98\cdot 10^{-6},3)$ & $(4.51\cdot 10^{-2},7)$ & $(0.444,4)$ & $4.67$ & $(6.07\cdot 10^{-2},2)$ & $4$      \\[1mm] \hline
$\tau_{ep}$ & $(3.36\cdot 10^{-6},1)$ & $(6.84\cdot 10^{-2},5)$ & $(0.545,3)$ & $3$    & $(1.63\cdot 10^{-2},1)$ & $2.5$    \\[1mm] \hline
$R_{mv}$       & $(1.16\cdot 10^{-5},5)$ & $(5.02\cdot 10^{-2},6)$ & $(0.365,6)$ & $5.67$ & $(12.0,8)$           & $6.25$   \\[1mm] \hline
$Z_{ao}$       & $(5.29\cdot 10^{-6},4)$ & $(5.28\cdot 10^{-4},9)$ & $(0.364,8)$ & $7$    & $(1.00,5)$           & $6.5$    \\[1mm] \hline
$R_s$        & $(8.96\cdot 10^{-4},8)$ & $(9.44\cdot 10^{-2},4)$ & $(0.888,1)$ & $4.33$ & $^{\dag}(2.70,7)$    & $5$      \\[1mm] \hline
$C_{sa}$       & $(3.67\cdot 10^{-5},6)$ & $(0.101,3)$          & $(0.364,7)$ & $5.33$ & $(0.364,4)$          & $5$      \\[1mm] \hline
$C_{sv}$       & $(9.76\cdot 10^{-2},9)$ & $(1.59\cdot 10^{-2},8)$ & $(0.363,9)$ & $8.67$ & $(21.0,9)$           & $8.75$   \\[1mm] \hline
$E_{max}$   & $(1.30\cdot 10^{-4},7)$ & $(0.124,2)$          & $(0.393,5)$ & $4.67$ & $^{*}(1.29,6,3)$     & $5,4.25$ \\[1mm] \hline
$E_{min}$   & $(3.72\cdot 10^{-6},2)$ & $(0.866,1)$          & $(0.816,2)$ & $1.67$ & $(0.339,3)$          & $2$      \\[1mm] \hline
\end{tabular}
    \centering
\end{table}

From Table \ref{tbl:results} we could choose an optimal set of input parameters based only on sensitivity and orthogonality. From the importance column, we could choose an optimal set of input parameters (1-6) given we exclude any parameter which does not contribute more than $5\%$ influence on the outputs. As we assume that we can not extract the parameters' effects on the outputs, they can not be readily identified. This then means $\tau_{es}, C_{sv}$ and $Z_{ao}$ are considered to be the unidentifiable parameters. From the error column, we find that the least sensitive parameter $Z_{ao}$ exhibits an error of only $1\%$, this is $5th$ smallest error and has $4$th smallest variance of $5.29\times 10^{-6}$. $\tau_{es}$ which we have assumed to be unidentifiable, exhibits the $2$nd smallest error of $0.0607\%$ and the $3$rd smallest parameter variance of $3.98\times 10^{-6}$. From Figure \ref{fig: UKF_COV}, input parameters which are recovered with minimal error also exhibit a steady variance.  We note from the above, a parameter may not be sensitive, but this does not mean that the parameter is not identifiable. 

When considering the dynamical equations we showed that both $\tau_{es}$ and $Z_{ao}$ interact independently with all three measurements $P_{lv},P_{sa}$ and $V_{lv}$, which implies that sensitivity is not the determinant of input parameter identifiability. We note that a parameter's variance reaching a steady minimum and interacting with all measurements appears to indicate that a parameter will be identifiable.
 
The sensitivity of parameters is not a redundant concept however; $C_{sa}$ and $E_{max}$ exhibit the $6$th and $7$th highest variances of the parameter set, higher than the variance of $R_{mv}$, which was found with an error of $12.0\%$. From this, one may conjecture that $C_{sa}$ and $E_{max}$ may be unidentifiable. However, $E_{max}$ and $C_{sa}$ rank as the 2$nd$ and $3$rd most influential input parameters across the cycle. In Figure \ref{fig: UKF_COV}, we see $E_{max}$ takes almost 30 cycles to reach a steady variance. If we compare this to the worst performing parameter $C_{sv}$, which has an error of $21.0\%$, we observe that $C_{sv}$ struggles to move away from its initial variance. If we compare $C_{sv}$ to $E_{max}$ and $C_{sa}$, we notice, the main difference is $E_{max}$ and $C_{sa}$ have considerably more influence over the cycle. We also see from our analysis of the dynamical equations $C_{sa}$ directly controls the $P_{sa}$ measurement so we can expect this parameter to be accurately recovered. In addition, $E_{max}$ interacts with all 3 measurements independently, whereas $C_{sv}$ does not interact with any measurements independently. Therefore, given a parameter does not reach a steady minimal variance, it appears that, if a parameter is sensitive and interacts with all measurements independently this can aid the parameter's efficient recovery.
 
In this research, we have also demonstrated the vital importance that a range of data have on the accurate recovery of input parameters. An exemplar is the systemic resistance $R_{s}$, an important parameter which is often used as a bio-marker of pathophysiology \cite{trammel2020physiology}. $R_{s}$ has an error of $2.7\%$ - considerably larger than the other errors observed in this work. From Table \ref{tbl:results} we see $R_{s}$ ranks $4$th in sensitivity and $1$st for orthogonality, indicating good identifiability. We see that $R_{s}$ never acts independently and is only exposed to the arterial pressure measurement $P_{sa}$. From the results discussed above, exposing $R_{s}$ to an orthogonal measurement along with arterial pressure (i.e., volume/flow), one might expect this parameter to be recovered more efficiently than in the current case. 

While we have discussed how important it is for the variance to reach a minimal level - which is an indicator for identifiability - we have also demonstrated the importance of subject expertise. With variance of said input parameter being a dominating factor in whether the parameter is identifiable or not, we note that during the modelling phase if one can include subject experts within research discussions this can lead to more assurance as to what the ``true" value of an input parameter is, hence reducing the initial uncertainty prescribed to the input parameter set. If this is not possible we have shown that the recovery of input parameters is still largely more accurate than the estimated values that come from medical equipment.
 
The real-time implementation of the UKF is conducted through a single ventricle model; as the model grows with complexity, we expect that its ability to compute input parameters within real-time may decrease. However, real-time is only a constraint which is needed for a systems model to be implemented into a clinical workflow \cite{huberts2018needed}, so given we can implement a more complex model into Julia which returns results in real-time, we have bridged the first step needed to integrate physiological models into clinical workflows. Aside from this, researchers interested in systems biological now have a simple usable example in which the only things needing changing are the model, the initial conditions and the noise errors associated with the system.

\section{Conclusion}
Broadly, we have quantitatively assessed the capability of the UKF accurately to recover input parameter values and states of a single ventricle model of the CV system by ingesting continuous and discrete time data. We use the SciML framework within the Julia language to develop an open-source, novel, real-time implementation. Our approach is amenable to a clinical workflow, can recover personalised patient state and parameter values with minimal error and may be applied to any system model.
 Specifically, input parameter global sensitivity and orthogonality analysis have been performed to assess the relationship between parameter recovery of the UKF and the sensitive and orthogonal input parameters. We have analysed a simple DAE model which contains 9 input parameters and have generated realistic synthetic clinical measurements. 
 
Perhaps surprisingly, we show that there is not an injective relationship between input parameter sensitivity and our filter's ability to recover input parameters and while high parameter sensitivity can contribute to accurate recovery of input parameters, it is not a sufficient condition. Rather, we find that a parameter's variance reaching a steady minimum is a stronger predictor of accurate assignment, 
compared with input parameter sensitivity and orthogonality analysis. Further, our analysis exposes the importance of utilising a diverse range of experimental data,
which allow identifiable parameters to be precisely recovered. Our observations question the received wisdom of deducing identifiable input parameters - we have shown that finding the identifiable parameters based solely on sensitivity and orthogonality methods may lead to misleading conclusions. The central concern of recovering parameter values from experimental data seems to be answered by the UKF which presents itself 
as a very practical method of uniquely identifying input parameters with low sensitivity. Our method is computationally inexpensive so the common approach of pre-screening input parameters to fix the unidentifiable ones is unnecessary. 
 
In the future, our method could be adapted to more realistic physiological models and applied to real clinical data, opening up new possibilities for personalised patient care. Directions for future research involve assessing a parameter's ability to adapt to perturbations within the cardiac cycle and adapting the method to deal with cycle-to-cycle heart rate variability. Exploring the relationship between input parameter sensitivity and the UKF's ability to recover parameters apparently has important, 
if nuanced practical implications for all system model parameter identification.  

\section*{Acknowledgments}
Harry Saxton is supported by a Sheffield Hallam University Graduate Teaching Assistant PhD scholarship.

% \section*{Appendices}
\appendix

\section{Model Derivation and Model Solution} \label{app-modelderiv}

In generic form, the equations relating to the passive compartmental 
state variables all take the form: 

\begin{align}
 \frac{dV_{s,i}}{dt} = Q_{i} - Q_{i+1}, \quad  \frac{dP_{i}}{dt} = \frac{1}{C_{i}}(Q_{i} - Q_{i+1}), \quad Q_{i} = \frac{P_{i}-P_{i+1}}{R_{i}}.
\end{align}

Above, the subscripts $(i-1),i,(i+1)$ respectively represent the proximal, present and distal system compartments, $V_{s,i}$(mL) denotes the circulating (stressed) volume  \cite{suga1974instantaneous} and $C_{i}$ (ml/mmHg) and $R_{i}$ (mmHgs/mL) denote  compartmental compliance and the Ohmic resistance between compartments $i$, $(i+1)$. See Figure \ref{fig:model} and Table \ref{tbl:measurements}. 
 
We return to system compartment number $1$ and its activation function shortly. 
Figure \ref{fig:model} is a schematic representation. 
Note, (i) we use a C-R-C Windkessel \cite{westerhof2009arterial} to represent the systemic circulation, (ii) no inertance appears in our formulation, (iii) the systemic and venous compartments are passive, having fixed compliance $C_{sa}$ and $C_{sv}$ respectively and (iv)  
flow in and out of the active left ventricle is controlled by the mitral and aortic valves respectively,  the latter being modelled as diodes, with Ohmic resistance under forward bias and infinite resistance under reverse bias
\begin{equation}
Q_{i}=
    \begin{cases}
        \frac{P_{i} - P_{i+1}}{R_{val}}, & P_i>P_{i+1}, \\
        0 & P_i \leq P_{i+1},
    \end{cases}
\end{equation}
where $R_{val}$ represents the resistance across the respective valves.
  
Let us consider the active model compartment. 
The dynamics of the left ventricle is described by a time varying compliance 
$C_{lv}(t)$, or reciprocal elastance, $E(t)$ (mmHg/ml) which determines the change in pressure for a 
given change in the volume \cite{suga1974instantaneous} 
\begin{equation}
\label{elastance}
    E(t) = \frac{P_{lv}(t)}{ V(t) - V_{0} } =  \frac{P_{lv}(t)}{V_{s}(t)},
\end{equation}
where $V_{0}$ \& $V_{s}(t)$ represent the unstressed and stressed volumes, respectively, in the chamber.
  
$E(t)$ may be described in analytical form as follows: \cite{korakianitis2006numerical}
\begin{align}
    E(t) & = (E_{max} - E_{min}) \cdot e(t) + E_{min}, \\
e(t)&=
    \begin{cases}
        \cos(\frac{\pi t}{\tau_{es}}) & 0 \leq t < \tau_{es}, \\
        \cos(\frac{\pi(t + \tau_{ep} - \tau_{es})}{\tau_{es}}) & \tau_{es}, \leq t < \tau_{ep}, \\
        0, & \tau_{ep} \leq t < \tau.
    \end{cases}
\end{align}
Above, $e(t;\tau_{es}, \tau_{ep})$ is the activation function, which 
is parameterised by the end systolic and end pulse timing parameters $\tau_{es}$ and $\tau_{ep}$ respectively.
 
The elastance function is defined over one cardiac cycle, i.e time $\Bar{t}\in[0,\tau]$ with 
$\tau$ (s) the length of the cardiac cycle, and contractility, $E_{max}$, and compliance, $E_{min}$, both of which control the left ventricular elastance extrema. There is a discontinuity in $E(t)$ at $t=\tau$.

\begin{figure}[htp!]
\centerline{\includegraphics[width=\columnwidth]{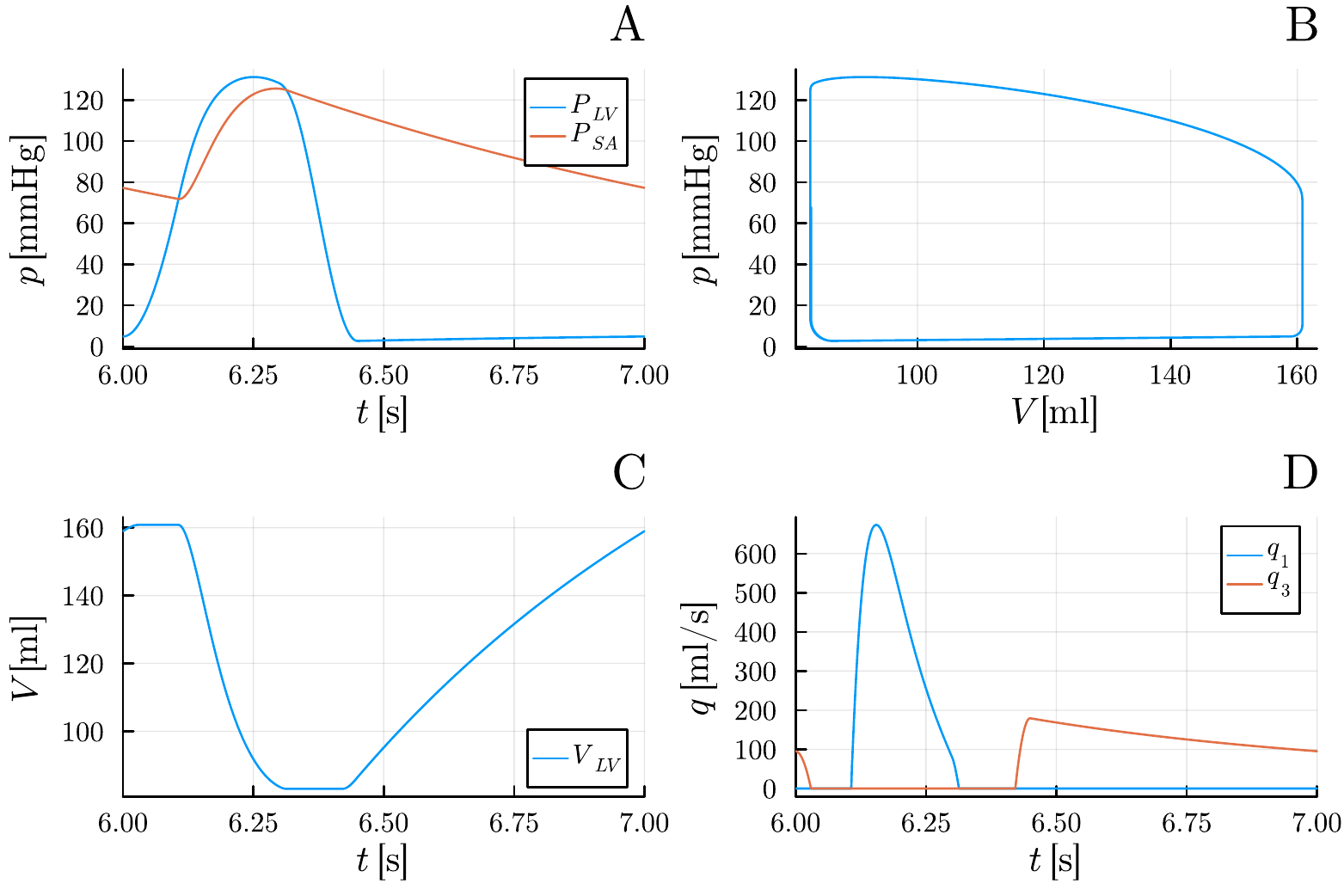}}
\caption{{ \bf Sample model solutions.} Shown are the model states of (A) ventricular and aortic pressures, (B) the PV-loop, (C) the ventricular volume and (D) the aortic and venous flow rates.}
\label{fig: Model_simulation}
\end{figure}

\section{Continuous Sensitivity Analysis}\label{app-ConSen}
\begin{figure}[htp!]
\centerline{\includegraphics[width=\columnwidth]{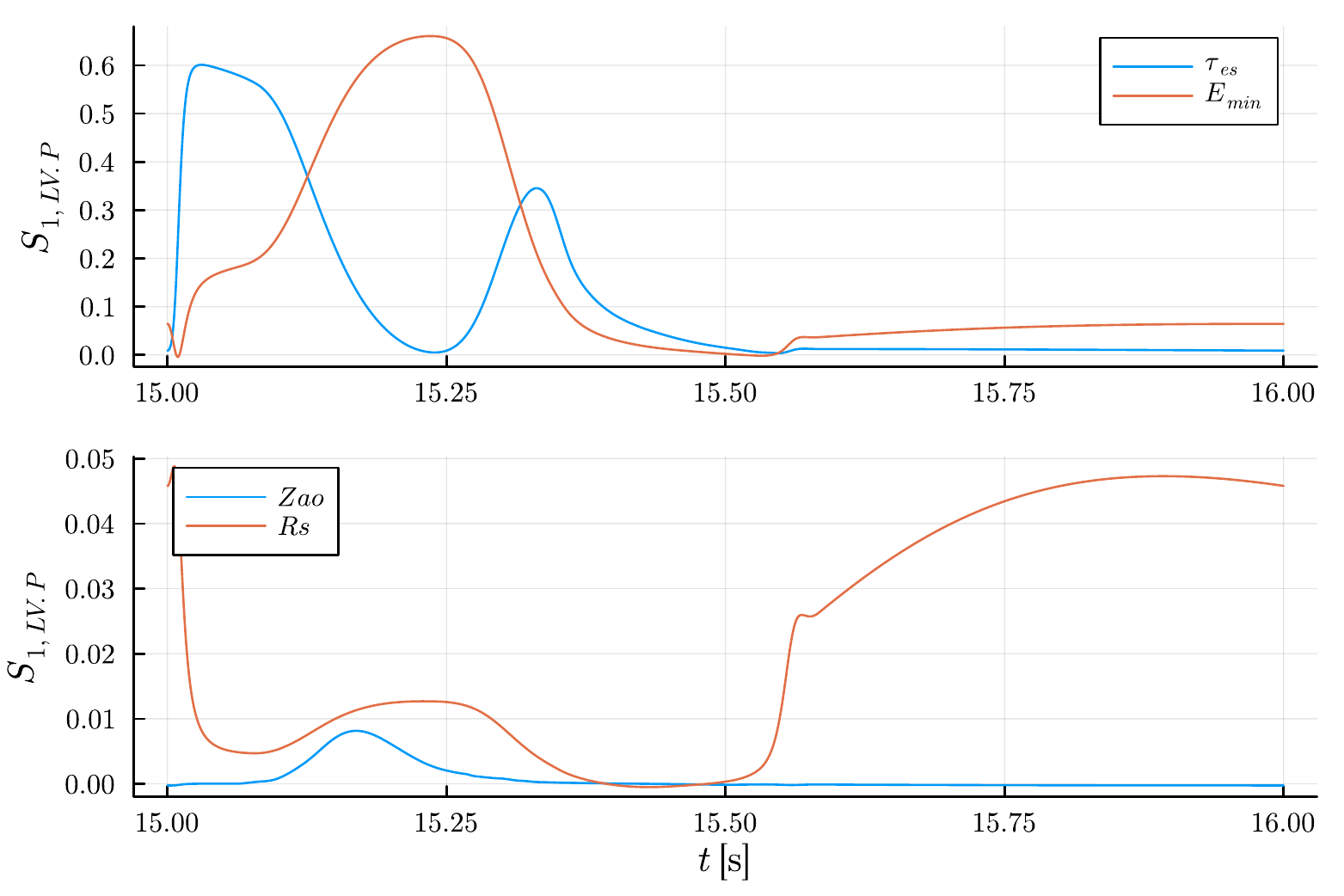}}
\caption{{ \bf Continuous Sensitivity Analysis.} Shown are the time-dependent first order Sobol indices for the ventricular volume $LV.P$ for both sensitive (top) and insensitive (bottom) parameters.}
\label{fig: Cont SA}
\end{figure}

From Figure \ref{fig: Cont SA} this displays the continuous waveform results for $\tau_{es}, E_{min}, Z_{ao}$ and $R_{s}$ we see that $Z_{ao}$ which displayed negligible sensitivity when averaged across the cycle exhibits some sensitivity at a point in the cycle which the UKF is able to get hold of. $\tau_{es}$ while not overly sensitive during the whole cycle is exhibits points during the cycle in which it displays large sensitivity which is why it is found with the second smallest error in the UKF estimation.

\section{Parameter Importance PCA Method} \label{app-ParamImpo}
We determine parameter influence, or effect, using principal component analysis (PCA) on equation (\ref{Sobol}), \cite{li2004selection,wold1987principal}. The PCs are the eigenvectors of a $n \times n$ Fisher information matrix (FIM) \cite{olsen2019parameter,fisher1922mathematical} based, note, upon first order Sobol indices: 
\begin{equation}
    \begin{aligned}
    \label{FIM}
    F=S_{1}^{T}S_{1}.
    \end{aligned}
\end{equation}
Let $Q$ be the matrix of ordered PC eigenvectors of $F$, in which the absolute value of each element $Q_{ij}$ reflects the contribution of the $jth$ parameter to the variance of the i$^{th}$ output. We follow Li et al. \cite{li2004selection}, who measure an overall effect for parameter j$^{th}$ as: 
\begin{equation}
    \begin{aligned}
    \label{overall_effect}
    e_{j} = \frac{\sum_{i=1}^{m}|\mu_{i}Q_{ij}|}{\sum_{i=1}^{m}|\mu_{i}|}.
    \end{aligned}
\end{equation}
Above, $\mu_{i}$ represents the non-zero eigenvalues of $F$.

\section{Analytical Work}\label{app-analytical}
During Systole, the equations are as follows, for the pressures, we also have the auxiliary equation in volume as a consequence of the relationship $E(t)=\frac{P_{lv}(t)}{V_{lv}(t)}$.
\begin{align*}
    \frac{dP_{lv}}{dt} &= -Q_{1}E(t) + \frac{P_{lv}}{E(t)}\frac{dE}{dt}, \hspace{1mm} \frac{dP_{sa}}{dt} = \frac{Q_{1}-Q_{2}}{Csa}, \\
    \frac{dP_{sv}}{dt} &= \frac{Q_{2}}{Csv}, \hspace{1mm} \frac{dV_{lv}}{dt} = -Q_{1},
\end{align*}
and the corresponding flows 
\begin{align*}
        Q_{1} = \frac{P_{lv} - P_{sa}}{Zao}, && Q_{2} = \frac{P_{sa} - P_{sv}}{Rs}. 
\end{align*}
No flows are measured in our synthetic dataset so we can eliminate flows which allow us to derive equations. Firstly we can examine the elastance relationship 
\begin{equation}
    E(t)=\frac{P_{lv}(t)}{V_{lv}(t)}.
\end{equation}
Here we see the elastance and its parameters are made up of two measured quantities hence it makes intuitive sense that one should be able to extract the parameters from the above relationship.
 
We examine the equation for the ventricular volume and find 
\begin{align}
    \frac{dV_{lv}}{dt} = \frac{P_{lv} - P_{sa}}{Z_{ao}}.
\end{align}
Here we note $Z_{ao}$ has direct influence on all the measured quantities. We can derive an equation for arterial pressure
\begin{align}
        \frac{dP_{sa}}{dt} = \frac{P_{lv} - P_{sa}}{C_{sa}Z_{ao}} - \frac{P_{sa} - P_{sv}}{C_{sa}R_s}. 
\end{align}
Here we note the effects $C_{sa}$ on the measured quantities are clouded by $Z_{ao}$ however due to $Z_{ao}$ acting independently in the ventricular volume equation the effects of $C_{a}$ can be quantified independently. In this equation we see that $R_{s}$ only interacts with arterial pressure and is clouded by the effects of $C_{sa}$. $C_{sa}$ directly controls our measured pressure $P_{sa}$.

\begin{align*}
    \frac{dP_{sv}}{dt} = \frac{P_{sa} - P_{sv}}{C_{sv}R_s}.
\end{align*}
From the venous pressure equation we see that the effects of $C_{sv}$ and $R_{s}$ are seen collectively on the measurement $P_{SA}$ and $R_{mv}$ on $P_{lv}$.
 
During Diastole we assume $Q_{1}=0$ the equations as: 
\begin{align*}
        \frac{dP_{lv}}{dt} &= Q_{3}E(t) + \frac{P_{lv}}{E(t)}\frac{dE}{dt}, \hspace{1mm} \frac{dP_{sv}}{dt} =\frac{Q_{2}-Q_{3}}{C_{sv}}, \\ 
        \frac{dP_{sa}}{dt} &= \frac{-Q_{s}}{C_{sa}}, \hspace{1mm}
        Q_{3} = \frac{P_{sv} - P_{LV}}{R_{mv}}, \hspace{1mm} Q_{2} = \frac{P_{sa} - P_{sv}}{R_s}.   \\
\end{align*}
Eliminating flow: 
\begin{align*}
        \frac{dP_{lv}}{dt}=  (\frac{P_{sv} - P_{LV}}{R_{mv}}&)E(t) + \frac{P_{lv}}{E(t)}\frac{dE}{dt}, \hspace{1mm}  \frac{dP_{sa}}{dt} = \frac{P_{sv} - P_{sa}}{C_{sa}R_s}, \\
        \frac{dP_{sv}}{dt} &= \frac{P_{lv}-P_{sv}}{C_{sv}R_{mv}} + \frac{P_{sa} - P_{sv}}{C_{sa}R_s}.
\end{align*}
Here we see that $R_{mv}$ interacts with $E(t)$ on $P_{lv}$ and $C_{sv}$ acts collectively with $R_{mv}$ on $P_{lv}$.


\bibliographystyle{splncs04}
\bibliography{Main_Paper}

\end{document}